\documentclass{article}
\usepackage{epsfig}

\date{\today}

\usepackage{amsmath}
\usepackage{amssymb}
\usepackage[hang,nooneline,scriptsize]{subfigure}
\begin{document}

\title{Stability of charged solitons and formation of boson stars in 5-dimensional Anti-de Sitter space-time}
\author{{\large Yves Brihaye \footnote{email: yves.brihaye@umons.ac.be}
}$^{(1)}$, 
{\large Betti Hartmann \footnote{email:
b.hartmann@jacobs-university.de}}$^{(2)}$
and
{\large Sardor Tojiev \footnote{email:
s.tojiev@jacobs-university.de}}$^{(2)}$
\\ \\
$^{(1)}${\small Physique-Math\'ematique, Universite de
Mons-Hainaut, 7000 Mons, Belgium}\\ 
$^{(2)}${\small School of Engineering and Science, Jacobs University Bremen,
28759 Bremen, Germany}  }

\date{\today}
\setlength{\footnotesep}{0.5\footnotesep}
\newcommand{\dd}{\mbox{d}}
\newcommand{\tr}{\mbox{tr}}
\newcommand{\la}{\lambda}
\newcommand{\ka}{\kappa}
\newcommand{\f}{\phi}
\newcommand{\vf}{\varphi}
\newcommand{\F}{\Phi}
\newcommand{\al}{\alpha}
\newcommand{\ga}{\gamma}
\newcommand{\de}{\delta}
\newcommand{\si}{\sigma}
\newcommand{\bomega}{\mbox{\boldmath $\omega$}}
\newcommand{\bsi}{\mbox{\boldmath $\sigma$}}
\newcommand{\bchi}{\mbox{\boldmath $\chi$}}
\newcommand{\bal}{\mbox{\boldmath $\alpha$}}
\newcommand{\bpsi}{\mbox{\boldmath $\psi$}}
\newcommand{\brho}{\mbox{\boldmath $\varrho$}}
\newcommand{\beps}{\mbox{\boldmath $\varepsilon$}}
\newcommand{\bxi}{\mbox{\boldmath $\xi$}}
\newcommand{\bbeta}{\mbox{\boldmath $\beta$}}
\newcommand{\ee}{\end{equation}}
\newcommand{\eea}{\end{eqnarray}}
\newcommand{\be}{\begin{equation}}
\newcommand{\bea}{\begin{eqnarray}}

\newcommand{\ii}{\mbox{i}}
\newcommand{\e}{\mbox{e}}
\newcommand{\pa}{\partial}
\newcommand{\Om}{\Omega}
\newcommand{\vep}{\varepsilon}
\newcommand{\bfph}{{\bf \phi}}
\newcommand{\lm}{\lambda}
\def\theequation{\arabic{equation}}
\renewcommand{\thefootnote}{\fnsymbol{footnote}}
\newcommand{\re}[1]{(\ref{#1})}
\newcommand{\R}{{\rm I \hspace{-0.52ex} R}}
\newcommand{\N}{{\sf N\hspace*{-1.0ex}\rule{0.15ex}%
{1.3ex}\hspace*{1.0ex}}}
\newcommand{\Q}{{\sf Q\hspace*{-1.1ex}\rule{0.15ex}%
{1.5ex}\hspace*{1.1ex}}}
\newcommand{\C}{{\sf C\hspace*{-0.9ex}\rule{0.15ex}%
{1.3ex}\hspace*{0.9ex}}}
\newcommand{\eins}{1\hspace{-0.56ex}{\rm I}}
\renewcommand{\thefootnote}{\arabic{footnote}}
 \maketitle
\begin{abstract} 
We study the stability of charged solitons in 5-dimensional Anti-de Sitter (AdS) space-time. We show that for
appropriate choices of the parameters of the model these solutions become unstable to form
scalar hair.  We find that the existence
of charged solitons with scalar hair depends crucially on the charge and the mass of the scalar field.
We investigate the dependence of the spectrum of solutions on the mass of the scalar field in detail.
For positive mass of the scalar field the hairy solitons can be interpreted as charged boson stars.
We find that for sufficiently small value of the charge of the scalar field a ``forbidden band''
of the boson star mass and charge exists, while all our results indicate that - contrary to the asymptotically
flat space-time case - boson stars in asymptotically AdS can have arbitrarily large charge and mass. 
\end{abstract}

\section{Introduction}
The Anti-de Sitter (AdS)/Conformal Field theory (CFT) correspondence relates gravity theories
in $(d+1)$-dimensional asymptotically AdS space-time to a CFT living on the $d$-dimensional boundary of that
space-time \cite{ggdual,adscft}. As such, classical solutions in asymptotically AdS have gained
a lot of interest. This includes both black hole solutions as well as globally
regular, solitonic-like solutions. Within the context of the holographic description
of high-temperature superconductivity solutions with asymptotic planar AdS have been considered 
\cite{gubser,hhh,horowitz_roberts,reviews}. It was shown that the formation of scalar (or vector) ``hair''
on the solutions is the dual description of the onset of superconductivity. This is possible
since the effective mass of the scalar field drops below the Breitenlohner-Freedman (BF) bound \cite{bf}
under certain circumstances and the black holes or solitons, respectively, become unstable to the 
formation of scalar hair possessing however still an asymptotic AdS space-time.
Besides being interesting due to the holographic interpretation, the stability of
classical field theory solutions is, of course, also of interest by itself as it could shed well light on
other interesting questions, e.g. black hole uniqueness. Consequently, the stability of black holes and solitons in global
AdS as well as in asymptotic hyperbolic AdS have also been discussed.
   
In \cite{Dias:2010ma} uncharged black holes in $(4+1)$ dimensions have been considered. It was shown that
static black holes with hyperbolic
horizons can become unstable to the formation of uncharged scalar hair on the horizon of the black hole
due to the existence of an extremal limit with near-horizon geometry AdS$_2\times H^3$ \cite{Robinson:1959ev,Bertotti:1959pf,Bardeen:1999px}. 
These studies were extended to include higher order curvature corrections in the form of 
Gauss-Bonnet terms \cite{hartmann_brihaye3}.

Charged black hole and soliton solutions in asymptotically global AdS in $(3+1)$ dimensions were studied in \cite{menagerie}.
It was pointed out  that the solutions tend to their planar counterparts for large charges since 
in that case the solutions can become comparable in size to the AdS radius. The influence of the Gauss-Bonnet
corrections on the instability of these solutions has been discussed in \cite{Brihaye:2012cb}. 
The corresponding investigation in $(4+1)$-dimensional global AdS was done in \cite{dias2,basu}. The existence
of solitons in this case had been suggested previously in a perturbative approach \cite{basu2}. In all studies
the mass of the scalar field had been set to zero. In \cite{brihaye_hartmannNEW} these results were extended 
to a tachyonic scalar field as well as to the rotating case. Recently, charged soliton solutions with positive
scalar field mass have been studied in $(3+1)$-dimensional global AdS \cite{hu}. In this case, the solutions carrying scalar
hair can be interpreted as charged
non-spinning boson star solutions in global AdS space-time. Uncharged boson stars in AdS were first discussed in 
$(d+1)$-dimensional AdS space-time using a massive
scalar field without self-interaction \cite{radu} and with an exponential self-interaction
potential \cite{hartmann_riedel,hartmann_riedel2}, respectively. Spinning solutions in $(2+1)$ and $(3+1)$ dimensions
have been constructed in \cite{radu_aste} and \cite{radu_subagyo}, respectively.

In this paper, we are interested in the stability of charged solitons in $(4+1)$-dimensional global AdS with respect
to formation of scalar hair. We investigate in detail the dependence of the pattern of solutions on the
scalar field mass and charge considering negative and positive values for the mass alike.
The work extends the analysis of \cite{dias2} and of \cite{bhs} to now include positive values for the mass
such that the solutions with scalar hair in our case can be interpreted as charged boson stars in $(4+1)$-dimensional
global AdS. 

This paper is organized as follow: in Section 2, we present the model, give the ansatz, the equations of motion and
the relevant boundary conditions.  It is well known that the generic (hairy) solutions cannot be constructed in
terms of elementary functions and that numerical techniques are needed. The discussion of the numerical
solutions is given in Section 3. Finally, Section 4 contains the conclusions.

\section{The model}

In this paper, we are studying the formation of scalar hair on electrically charged  solitons in 
$(4+1)$-dimensional Anti--de Sitter space--time. 
The action reads~:
\begin{equation}
S= \frac{1}{16\pi G} \int d^5 x \sqrt{-g} \left(R -2\Lambda  + 16\pi G {\cal L}_{\rm matter}\right) \ ,
\end{equation}
where $\Lambda=-6/L^2$ is the cosmological constant, $R$ denotes the Ricci scalar and $G$ is Newton's constant.
${\cal L}_{\rm matter}$ denotes the matter Lagrangian which reads~:
\begin{equation}
{\cal L}_{\rm matter}= -\frac{1}{4} F_{MN} F^{MN} - 
\left(D_M\psi\right)^* D^M \psi - m^2 \psi^*\psi  \ \ , \ \  M,N=0,1,2,3,4  \ ,
\end{equation}
where $F_{MN} =\partial_M A_N - \partial_N A_M$ is the field strength tensor and
$D_M\psi=\partial_M \psi - ie A_M \psi$ is the covariant derivative.
$e$ and $m^2$ denote the electric charge and mass of the scalar field $\psi$, respectively.

We choose the following spherically symmetric Ansatz for the metric~:
\begin{equation}
ds^2 = - f(r) a^2(r) dt^2 + \frac{1}{f(r)} dr^2 + \frac{r^2}{L^2} d\Omega^2_{3}  \ ,
\end{equation}
where $f$ and $a$ are functions of $r$ only and $d\Omega^2_3$ denotes the line element of a unit 3-dimensional
sphere. 

For the electromagnetic field and the scalar field we have \cite{hhh}~:
\begin{equation}
A_{M}dx^M = \phi(r) dt \  \  \  , \   \   \   \psi=\psi(r)
\end{equation}
such that the solutions possess only electric charge.

The coupled gravity and matter field equations are obtained from the variation of the
action with respect to the matter and metric fields, respectively, and read
\
\begin{eqnarray}
\label{eq1}
     f' &=& \frac{2}{r} \left(1-f+ 2 \frac{r^2}{L^2}\right)
     - \gamma \frac{r}{2 f a^2} 
     \left(2 e^2 \phi^2 \psi^2 + f (2 m^2 a^2 \psi^2 + \phi'^2) + 
2 f^2 a^2 \psi'^2\right)  \ , \\
\label{eq2}
        a' &=& \gamma \frac{r(e^2 \phi^2 \psi^2 + a^2 f^2 \psi'^2)}{a f^2}  \ ,  \\
\label{eq3}
   \phi'' &=& - \left(\frac{3}{r} - \frac{a'}{a}\right) \phi' +
2 \frac{e^2 \psi^2}{f} \phi  \ , \\
\label{eq4}
    \psi'' &=& -\left(\frac{3}{r} + \frac{f'}{f} + \frac{a'}{a}\right) \psi'- 
\left(\frac{e^2 \phi^2}{f^2 a^2} - \frac{m^2}{f}\right) \psi  \ ,
\end{eqnarray}
where $\gamma=16\pi G$. Here and in the following the prime denotes
the derivative with respect to $r$. 
These equations
depend on the following independent constants: Newton's constant $G$, 
the cosmological constant $\Lambda$ (or Anti-de Sitter radius $L$) and the charge $e$ and mass $m$ of the 
scalar field.

The system possesses two scaling symmetries:
\begin{equation}
\label{scaling1}
 r\rightarrow \lambda r \ \ \ , \ \ \ t\rightarrow \lambda t \ \ \ , \ \ \ L\rightarrow \lambda L
\ \ \ , \ \ \ e\rightarrow e/\lambda \ \ \ 
\end{equation}
as well as 
\begin{equation}
\label{scaling2}
 \phi \rightarrow \lambda \phi \ \ \ , \ \ \ \psi \rightarrow \lambda \psi \ \ \ , \ \ \ 
e\rightarrow e/\lambda \ \ \ ,  \ \ \ \gamma\rightarrow \gamma/\lambda^2  \ ,
\end{equation}
which we can use to set $L=1$ and $\gamma$ to some fixed value without loosing generality.

In order to solve the equations, we have to fix appropriate boundary conditions. 
For soliton solutions regular on $r \in [0, \infty[$  we require
\be
  f(0) = 1 \ \ , \ \ \phi'(0) = 0 \ \ , \ \ \psi(0) \equiv \psi_0 \ \ , \ \ \psi'(0) = 0  \ ,
\ee
where $\psi_0$ is a free parameter.

Asymptotically, we want the space--time to be that of global AdS, i.e. we can choose
$a(r\rightarrow\infty)\rightarrow 1$. Other choices of the asymptotic value
of $a(r)$ would simply correspond to a rescaling of the time coordinate. The matter fields on the other hand obey~:
\begin{equation} 
  \phi(r\gg 1) = \mu - \frac{Q}{r^2}  \ \ , \ \ 
  \psi(r\gg 1) = \frac{\psi_{-}}{r^{\lambda_{-}}} + \frac{\psi_{+}}{r^{\lambda_{+}}} \ \
\label{decay}
\end{equation}
with
\begin{equation}
\label{lambda}
       \lambda_{-} = 2 - \sqrt{4 +m^2 L^2} \ \ , \ \ \lambda_{+} = 
2 + \sqrt{4 + m^2 L^2} \ \  \ .
\end{equation}
The parameters $\mu$, $Q$ are the chemical potential and the electric charge, respectively. 
Note that for $m^2 \geq 0$ we have to choose $\psi_-=0$. We will hence also adapt this choice 
for $m^2 < 0$. 
Within the holographic interpretation of our model $\psi_{+}$ will correspond to the expectation 
value $\langle{\cal O}\rangle$ of the operator
${\cal O}$  dual
to the scalar field on the conformal boundary of AdS.
Note that in comparison to the case of holographic superconductors (see e.g. \cite{hhh}
and references therein)
the space-time in this study possesses spherically symmetric sections for constant $r$ and $t$.
In other words, for $r\rightarrow \infty$ our space-time corresponds to global
AdS. 

The mass of the solution is given in terms of the coefficients that appear in the
asymptotic fall-off of the metric functions which reads
\begin{equation}
  f(r\gg 1) =  1+ \frac{r^2}{L^2} + \frac{f_2}{r^2} + \dots \ \ , \ \ 
 a(r\gg 1) = 1 + \frac{c_4}{r^4} + \dots \ .
\end{equation}
$f_2$ and $c_4$ are constants that have to be determined numerically and that
depend on the couplings in the model.
The mass $M$ then reads
\begin{equation}
   M = - \frac{f_2}{2} \ .
\end{equation}

\section{Numerical results}
We have used a Newton-Raphson algorithm with adaptive grid scheme \cite{colsys} to construct the solutions numerically. 
In the numerical construction, we set $L^2=1$ and $\gamma = 9/40$ without loss of generality. 
The remaining parameters of the model are $e^2$ and $m^2$. 

Once fixing $e^2$, $m^2$ families of regular solutions labelled by $\psi(0)\equiv\psi_0$ (and then by $Q$)
can be constructed numerically. The pattern of solutions turns out to be so involved that none of these quantities 
can be used to parametrize the  branches  uniquely: in some regions two or more solutions are characterized 
by the same value of $Q$ or of $\psi_0$. Furthermore, for the values of the parameters studied we noticed
the occurrence of at least two disconnected branches. 
A similar phenomenon was observed in different models previously \cite{dias2, gentle}.

\subsection{$m^2=1$}
We  first discuss the solutions for positive values of the mass parameter $m^2$ choosing $m^2=1$. 
In this case the soliton solutions carrying scalar hair can be interpreted as charged boson stars in global AdS.

Fixing in addition $e^2$ to a particular value we have constructed a first 
branch of solutions which can be labelled by $\psi_0 \equiv \psi(0)$. Letting $\psi_0\to 0$ we find that
$\phi(r)=\psi(r)\equiv 0$, while $a(r) \equiv 1$. The resulting solution hence corresponds
to global AdS space-time. Accordingly, the mass $M$ and the charge $Q$ tend to zero in this limit.

\begin{figure}[h]
\begin{center}

\subfigure[][$M$ as function of $\psi(0)$]{\label{fig1a}
\includegraphics[width=6cm]{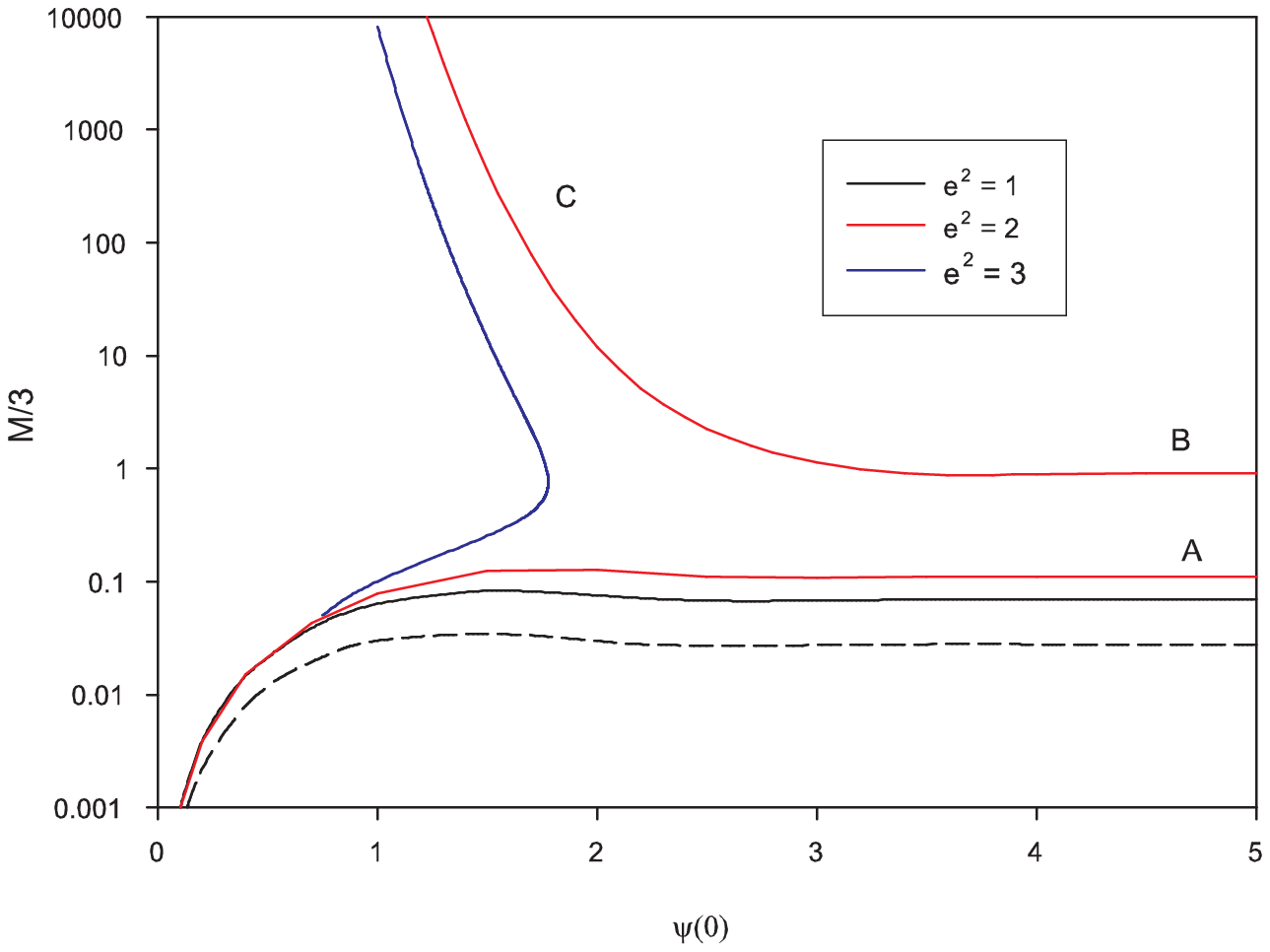}}
\subfigure[][$a(0)$ as function of $\psi(0)$  ]{\label{fig1b}
\includegraphics[width=6cm]{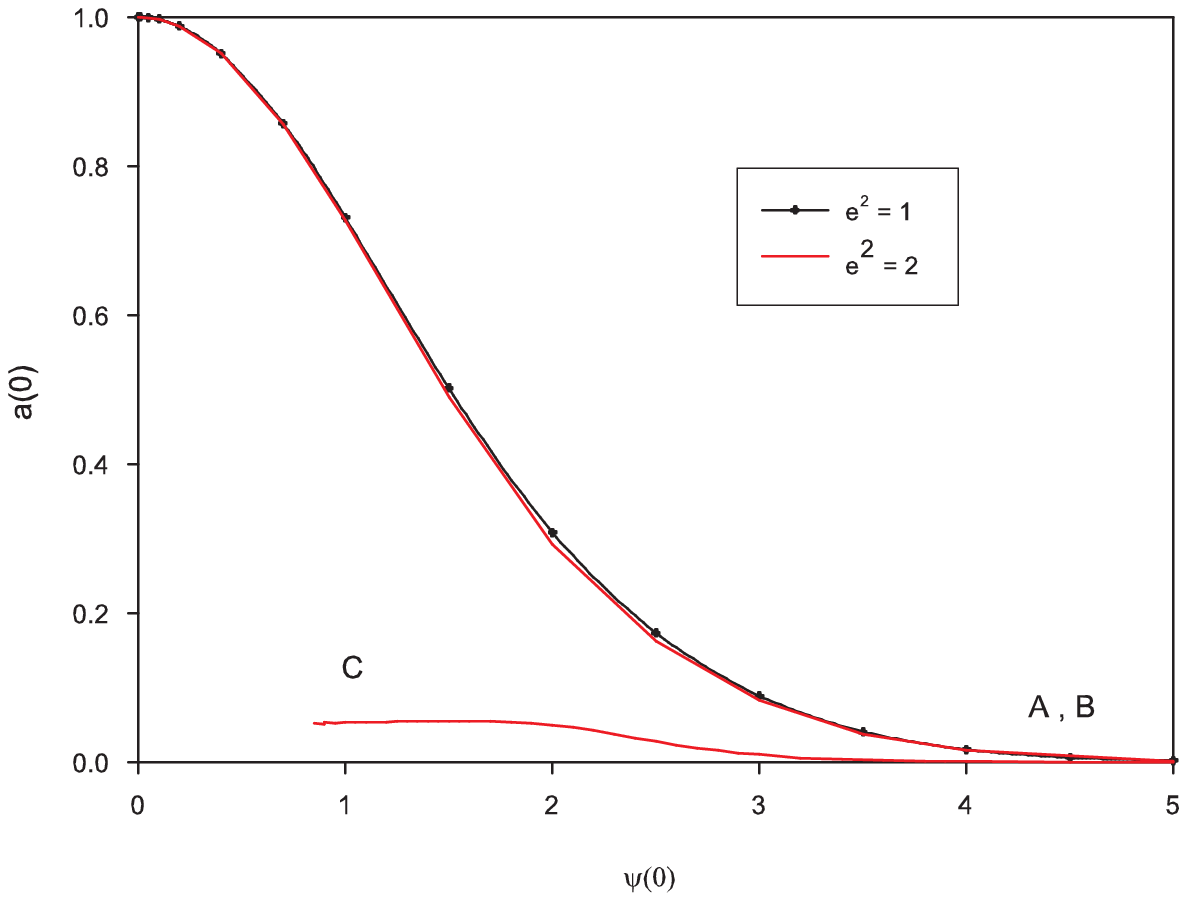}} \\
\subfigure[][$M$ as function of $Q$]{\label{fig1c}
\includegraphics[width=6cm]{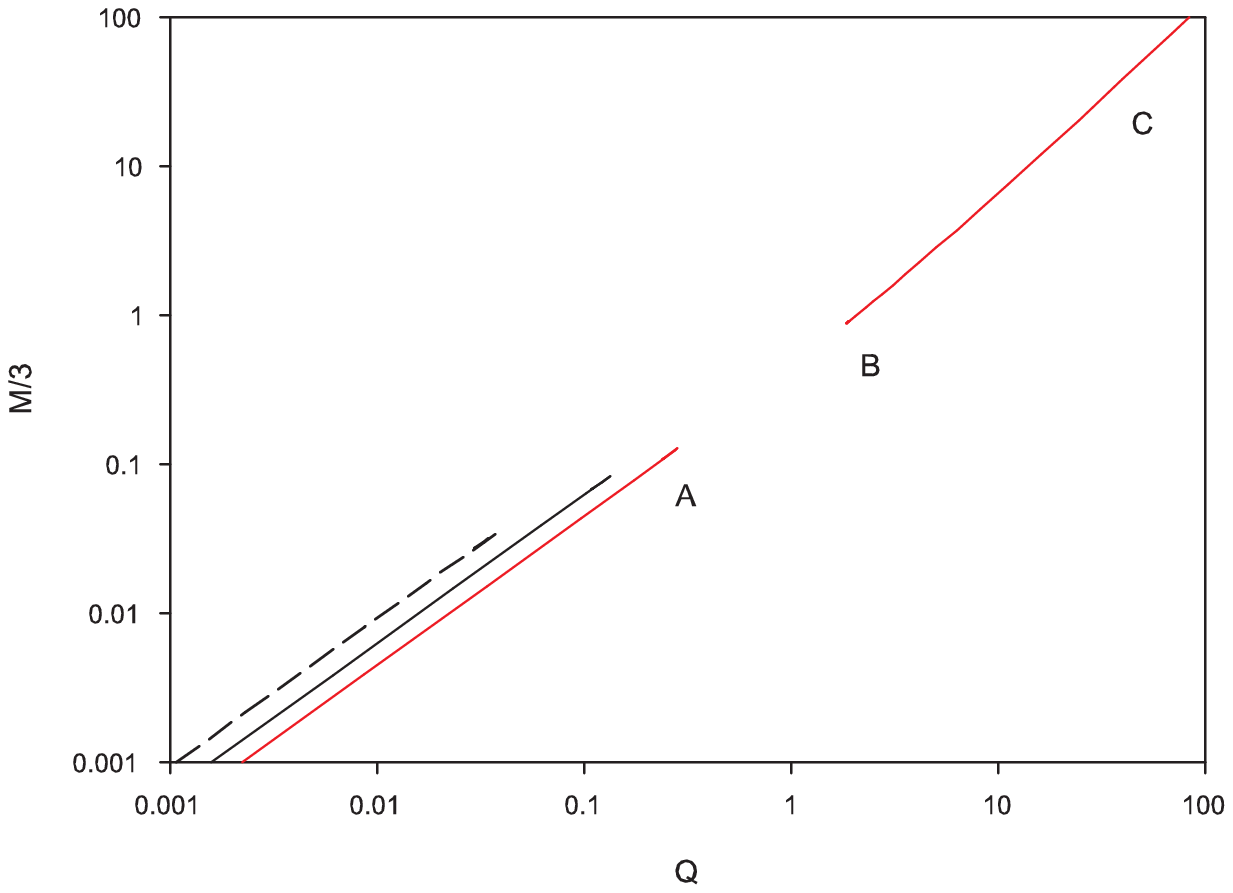}}
\subfigure[][$F=M-\mu Q$ as function of $Q$]{\label{fig1d}
\includegraphics[width=6cm]{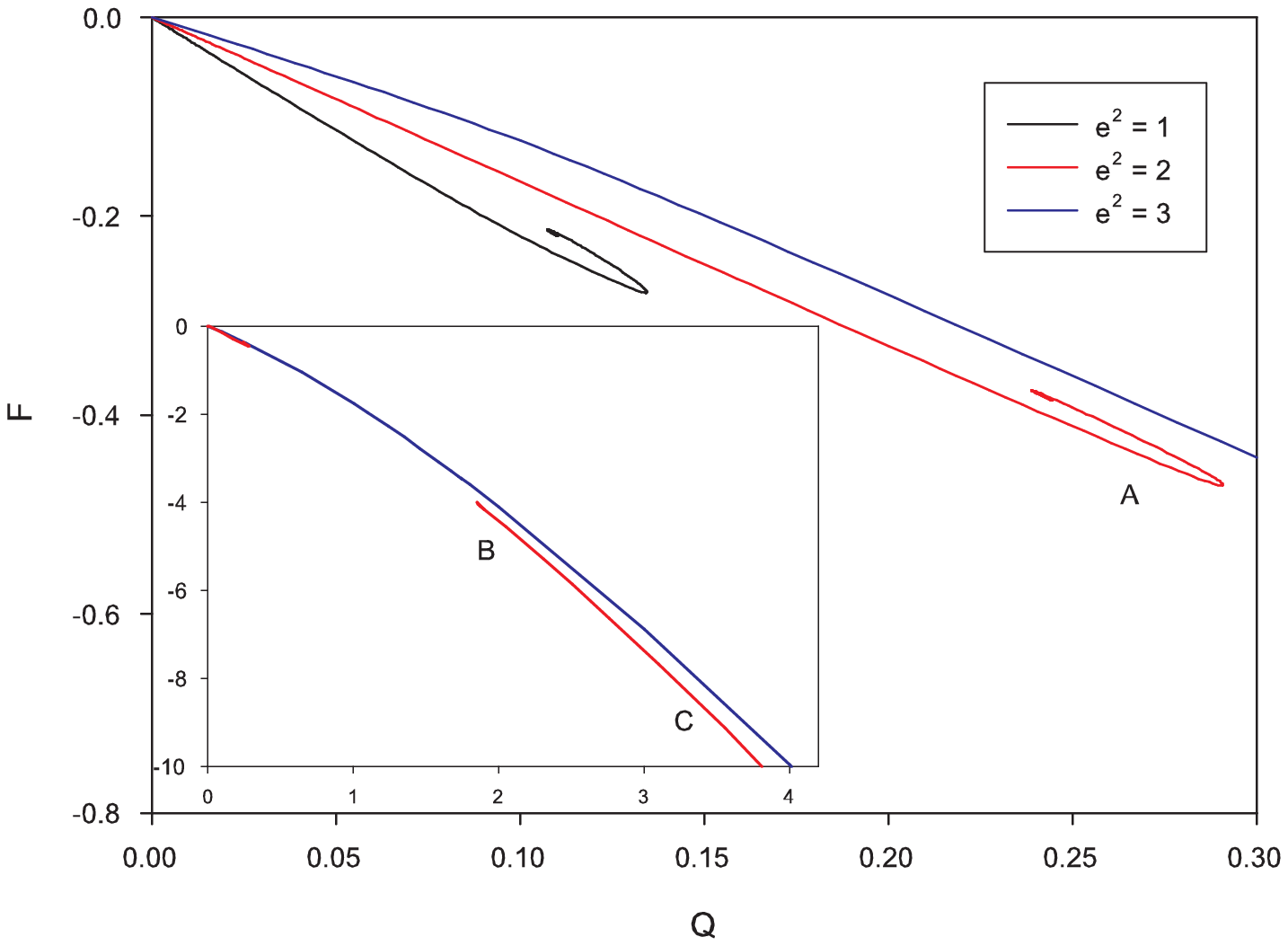}}
\end{center}
\caption{\label{fig_2} We show the mass $M$ (upper left) and the value of $a(0)$ 
(upper right) as functions of $\psi_0$ for 
$m^2=1$ and for $e^2=1$ (black), $e^2=2$ (red) and $e^2=3$ (blue), respectively. 
We also give the mass $M$ (lower left) as well as the free energy $F=M-\mu Q$ (lower right)
in dependence on $Q$.
Note that the middle and lower figure 
do not contain the results for $e^2=3$. This is due to the fact that the data for $e^2=3$ is so close
to that for $e^2=2$ that the curves cannot be distinguished. Furthermore, in the $M$-over-$Q$ plot the curve
for $e^2=3$ would be continuous and hence the gap for $e^2=2$ and $e^2=1$ would not be clearly visible. 
In all figures the solid lines represent the fundamental solutions, while the dashed lines
correspond to the first excited solutions (here given for $e^2=1$). }    
\end{figure}

We find that the solutions' properties are crucially related to
the choice of $e^2$. We first choose three specific value of $e^2$ to demonstrate the general pattern.
Our results are shown in Fig.\ref{fig_2}, where we give $M$ and $a(0)$ as functions of $\psi(0)\equiv \psi_0$ 
(upper left and right figure, respectively)
and also present $M$ as well as the free energy $F=M-\mu Q$ as functions of $Q$ (lower left and right figure, respectively). 
Note that Fig.\ref{fig1b} and Fig.\ref{fig1c} 
do not contain the results for $e^2=3$. This is due to the fact that the data for $e^2=3$ is so close
to that for $e^2=2$ that the curves cannot be distinguished. Furthermore, in the $M$-over-$Q$ plot the curve
for $e^2=3$ would be continuous and hence the gap for $e^2=2$ and $e^2=1$ would not be clearly visible. 
The solid lines correspond
to the fundamental solutions for which the scalar field function $\psi(r)$  
decreases monotonically  from  $\psi_0 > 0$ to zero at $r=\infty$. On the other hand the dashed line (for $e^2=1$) 
corresponds to first excited solutions, i.e.
solutions that possess one node in the scalar field function. 

Our results suggest the existence of a critical value $e^2_{\rm cr}$ of the parameter $e^2$ that separates the
pattern of solutions into (at least) two types. To be more precise, we find that 
for $e^2 < e_{\rm cr}^2$ two branches of solitons with scalar hair can be constructed. 
One branch - which we will call the ``main branch'' in the following - is connected to the AdS vacuum at $\psi(0)=0$ 
(and $M=Q=0$, $a(0)=1$) and can be continued to arbitrarily high values of $\psi_0$ 
(see label A in Fig.\ref{fig_2}). In this limit, the mass $M$ and $Q$ tend to finite values, while $a(0)$ tends
to zero. Note that in this case the so-called ``planar limit'' \cite{gentle} which appears 
for large $Q$ cannot be reached
by a smooth deformation of the AdS vacuum. Next to this main branch, a second branch of hairy soliton exists (label B and label C in Fig.\ref{fig_2}). This
is completely disjoint for the main branch. Again, solutions exist for arbitrarily high values of $\psi_0$. In this limit
(see label B) the mass $M$ and the charge $Q$ are again finite, however higher in value than the corresponding
values in the same limit on the main branch (see label A). All our numerical results indicate
that the two branches do {\it not} join for sufficiently large values of $\psi_0$, but remain more or less parallel.
This leads to the observation that a ``forbidden band'' for the mass $M$ and equivalently for the charge $Q$ exists
for charged boson stars
in asymptotically AdS if the charge of the scalar field $e^2$ is ``small enough''. 
The plot of the free energy $F=M-\mu Q$ tells us which solutions are stable for a fixed value of $Q$ (since we are
in the canonical ensemble). Fig.\ref{fig1d} clearly shows that close to the critical value of $Q$, where the main branch
of solutions terminates a spiralling can be observed. The solutions with the lowest free energy are the preferred ones, i.e.
the solutions that are directly connected to the AdS vacuum. Moreover, we observe that for a fixed value of $Q$ the value of
the free energy decreases with decreasing $e^2$. This is true for both the main as well as for the second branch.

In the limit $\psi_0\rightarrow 0$ the solutions on the second branch do not tend to the AdS vacuum, but
the mass $M$ and charge $Q$ tend to infinity (see label C). All these features are clearly visible in Fig.\ref{fig_2}
for $e^2=2$. For $e^2=1$ we also find the main branch and all our numerical results indicate that only this
branch exists. If the second branch also existed it would be extremely difficult to construct this numerically.

Another striking feature of the second branch of solutions
(see label C in Fig. \ref{fig_2}) is that it apparently extends backwards to $\psi_0=0$ 
although the numerical construction becomes difficult in this limit. 
A solution in this parameter range is given in Fig. \ref{fig_8}. We observe that 
the metric function $f(r)$ possesses a minimum at some intermediate value of the
radial coordinate $r=r_{\rm min}$ which increases with decreasing $\psi_0$. Furthermore, 
the scalar field $\psi(r)$ deviates only little from the value $\psi_0$  up to $r=r_{\rm min}$, while 
for $r>r_{\rm min}$ the scalar function reaches its power-like behaviour. 

\begin{figure}
\centering
\epsfysize=8cm
\mbox{\epsffile{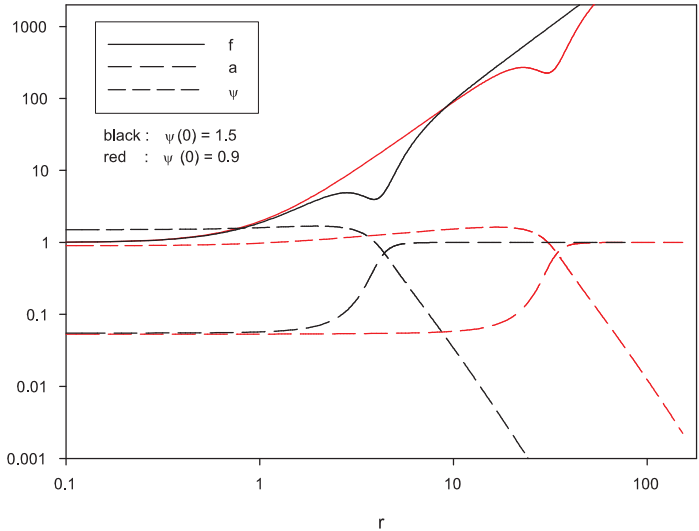}}
\caption{\label{fig_8}
We show the profiles of the metric functions $f(r)$ (solid) and $a(r)$ (long-dashed) as well as of the 
scalar field function 
$\psi(r)$ (short-dashed) for 
 $e^2=2$,  $m^2=1$ and for $\psi(0)= 1.5$ (black) and $\psi(0)=0.9$ (red), respectively.}
\end{figure}

Let us also remark that the qualitative features are similar for the first excited solutions, i.e. the solutions
that possess one zero in the scalar field function. This is shown for $e^2=1$ in Fig.\ref{fig_2} (dashed line).
The dependence of the mass $M$ on $\psi_0$ is similar than that of the fundamental solution and the $M$-$Q$-plot
also clearly shows that the mass of the excited solutions is larger than that of the fundamental solution
for a fixed value of $Q$. These solutions are hence unstable to decay to the fundamental ones. 

For $e^2 > e_{\rm cr}^2$ we find that only one branch of solutions exists. This branch can be constructed
from $\psi(0)=0$ (and $M=Q=0$) at the AdS vacuum up to a maximal finite value of $\psi(0)=\psi(0)_{\rm max}$.
From this maximal value of $\psi(0)$ the branch can be continued when decreasing $\psi(0)$ again. On this
back-bending branch the mass $M$ and charge $Q$ of the solution tends to infinity. Accordingly, the solutions tend to 
the ``planar limit'' \cite{gentle} in which the size of the solitons becomes comparable to the AdS radius. 
The shape of the solution on this branch 
is reminiscent of the so-called ``thick-wall limit'' \cite{kusenko} of uncharged $Q$-balls and boson stars
which appears for $\psi_0\rightarrow 0$. This also means that we can find a branch of solutions that interpolates
smoothly between the AdS vacuum and the planar limit. 
Finally, we find that for $e^2 = e_{\rm cr}^2$ the main branch and the second branch join at a specific 
value of $\psi(0)$. 

We have then done a detailed analysis of the solutions around the critical
value of $e^2$. Our results are given in Fig.\ref{fig_1}, where we show the mass $M$ as function of $\psi(0)$ for several
values of $e^2$ close to the critical value of $e^2$ which we determine to be $e_{\rm cr}^2\approx 2.6529$ for $m^2=1$. 
This should be compared to the value for $m^2=0$, which
is $e_{\rm cr}^2 \approx 2.3696$. Hence, $e_{\rm cr}^2$ increases with increasing $m^2$.
 
These results can be compared to the analogue case in $(3+1)$ dimensions \cite{hu}. 
\begin{figure}
\centering
\epsfysize=7cm
\mbox{\epsffile{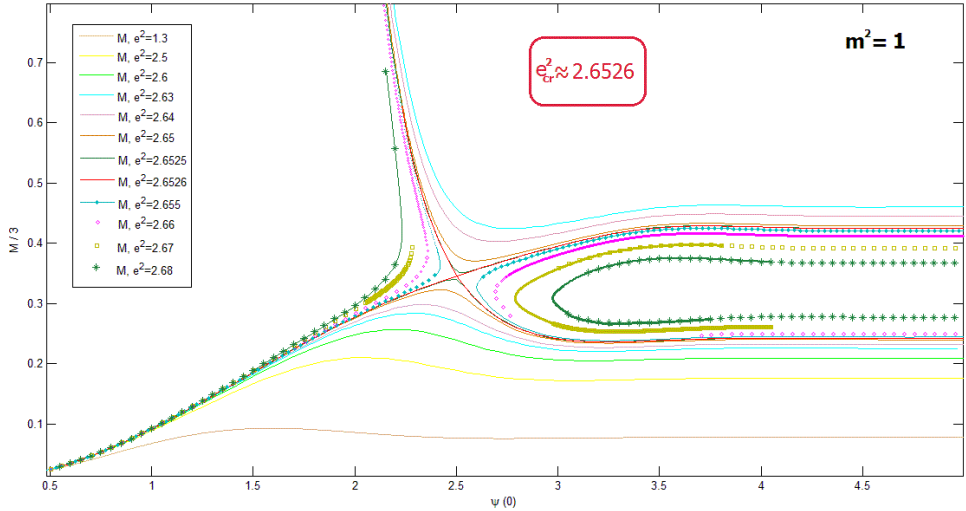}}
\caption{\label{fig_1}
 The mass $M$ is shown as functions of $\psi(0)\equiv\psi_0$ for 
$m^2=1$ and several values of $e^2$ close to the critical value $e_{\rm cr}^2$.}
\end{figure}

\subsection{$m^2=-3$}
If we want the asymptotic AdS of our space-time to be stable the mass $m^2$ of the 
scalar field
is limited by the BF bound (with our choice $L=1$ we have $m^2_{BF}=-4$ in $(4+1)$ dimensions).  
We have therefore studied  the spectrum of the hairy soliton solutions for a value of $m^2$
relatively close to the BF bound, namely $m^2=-3$. 

To appreciate the difference with respect to the case $m^2=1$ 
we present the counterpart of Fig.\ref{fig_2} in Fig.\ref{fig_2_2}.

\begin{figure}[h]
\begin{center}

\subfigure[][$M$ as function of $\psi(0)$]{\label{fig2a}
\includegraphics[width=6cm]{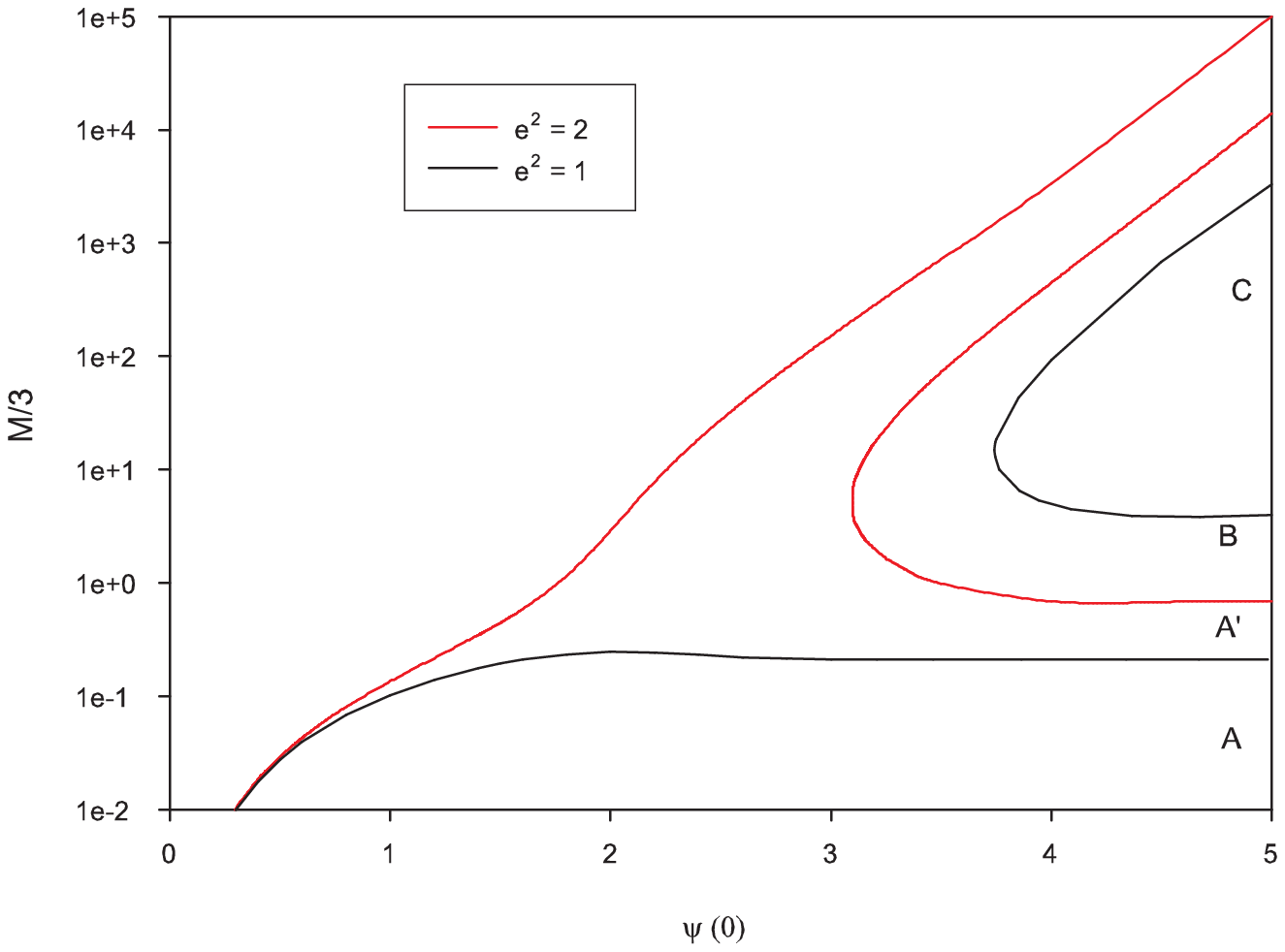}}
\subfigure[][$a(0)$ as function of $\psi(0)$  ]{\label{fig2b}
\includegraphics[width=6cm]{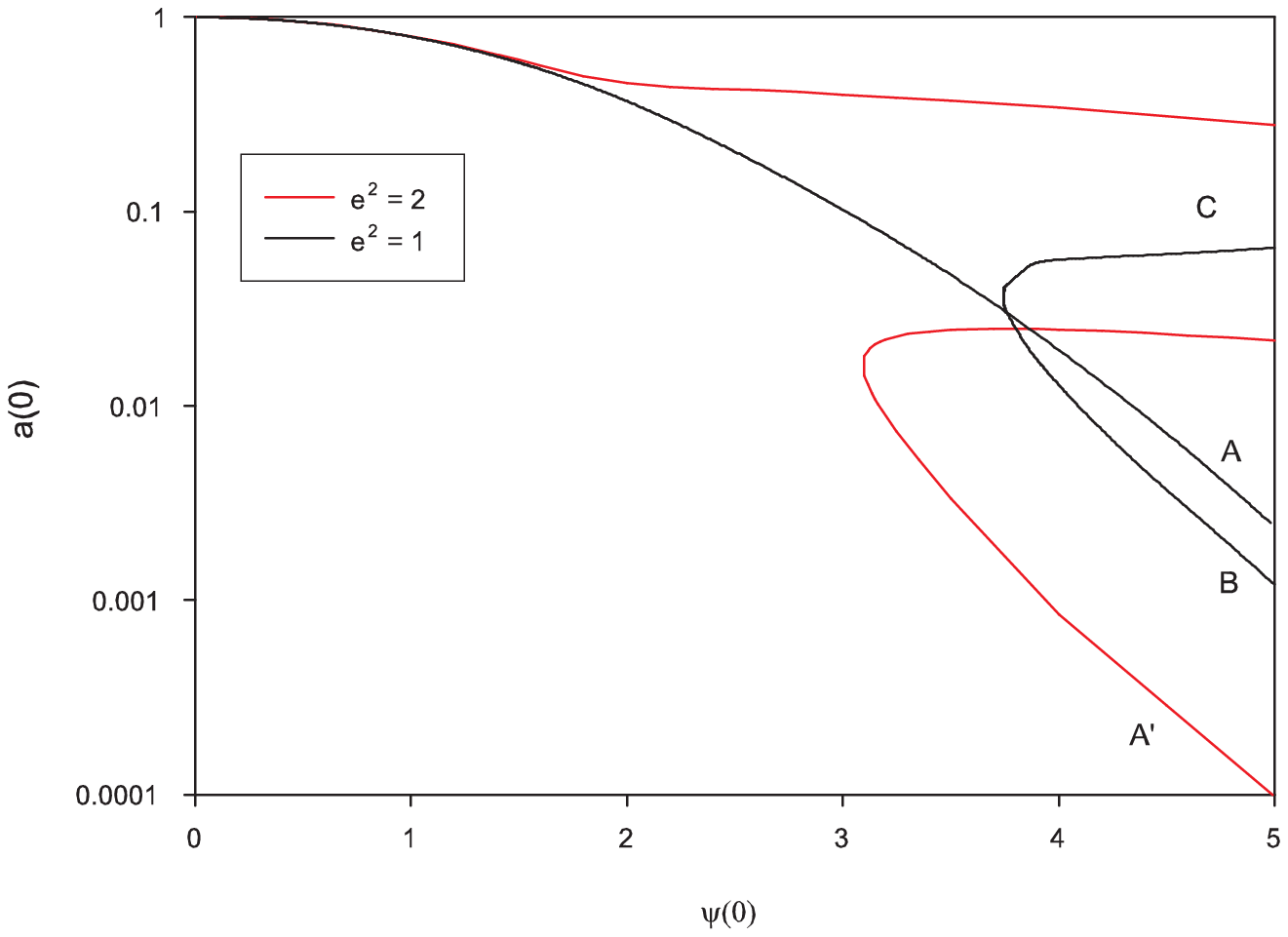}} \\
\subfigure[][$M$ as function of $Q$]{\label{fig2c}
\includegraphics[width=6cm]{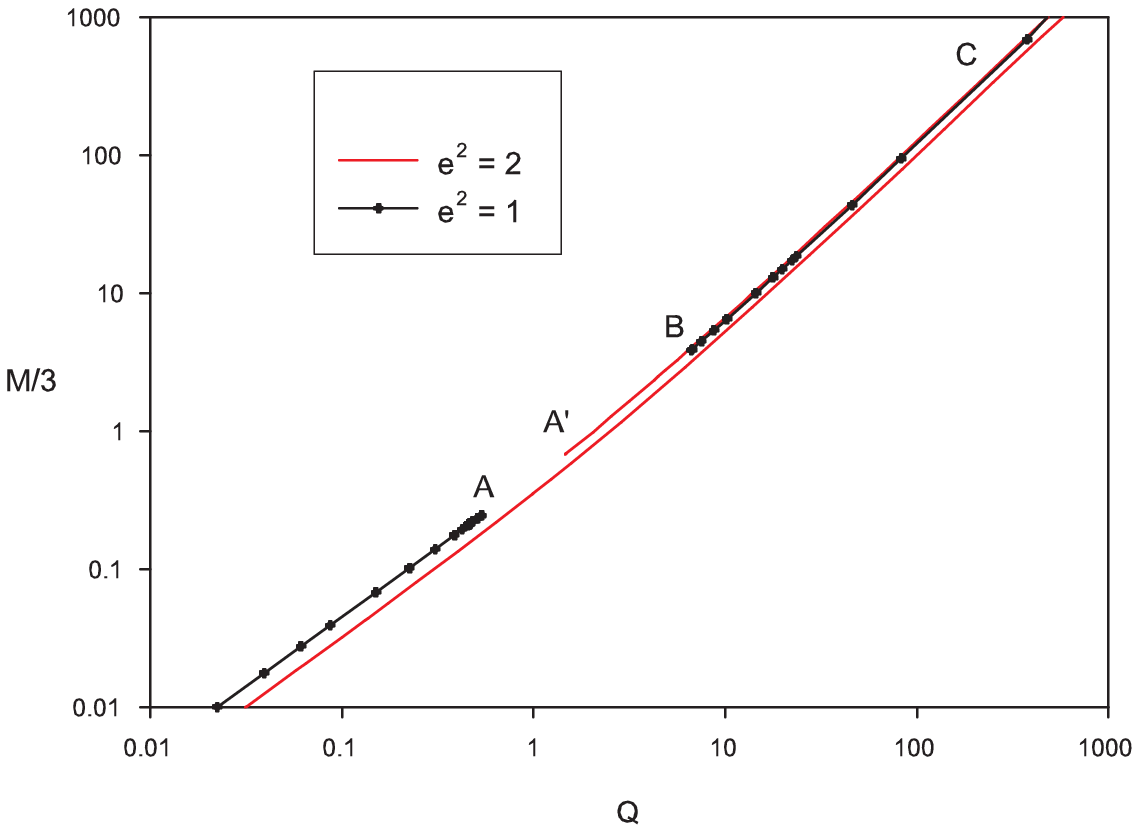}}
\subfigure[][$F=M-\mu Q$ as function of $Q$]{\label{fig2d}
\includegraphics[width=6cm]{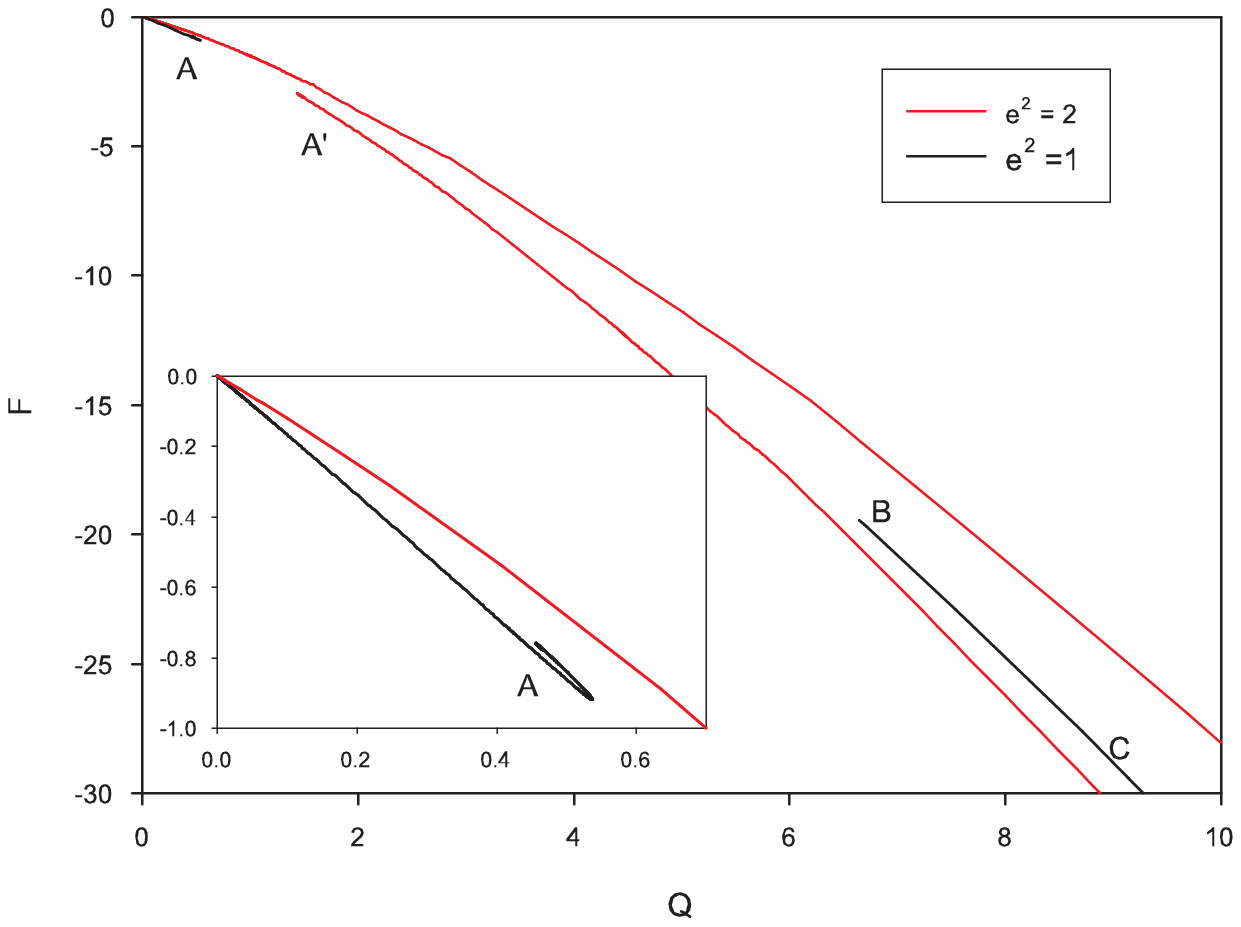}}
\end{center}
\caption{\label{fig_2_2} We show the mass $M$ (upper left) and the value of $a(0)$ (upper right) as 
functions of $\psi_0$ for 
$m^2=-3$ and for $e^2=1$ (black) and $e^2=2$ (red), respectively. We also give the mass $M$ (lower left)
as well as the free energy $F=M-\mu Q$ (lower right) in dependence on $Q$.
In all figures the labels A, B, C represent branches of fundamental solutions,
while A'
corresponds to a branch of first excited solutions (here given for $e^2=1$).}    
\end{figure}

Here, we find that there is again a critical value of $e^2$ that separates the pattern
into distinct types. Our numerical results indicate that $e_{\rm cr}^2 \approx 1.3575$. For $e^2 < e_{\rm cr}^2$
(here $e^2=1$, black curves) we again find a main branch of solutions which is connected to the AdS vacuum
at $\psi(0)=0$, $M=Q=0$. On this main branch the solutions can be constructed for arbitrarily large values 
of $\psi(0)$ and in the limit $\psi(0)\rightarrow \infty$ (label A in Fig.\ref{fig_2_2})
the mass $M$ and charge $Q$ tend to finite values, while $a(0)$ tends to zero. 
Similarly, a second branch of solutions exists in
this case, however, this is only present for $\psi(0)$ larger than a minimal value $\psi_{\rm min}(0)$ 
(label B). 
At this minimal value a
third branch of solutions can be constructed that extends back to $\psi(0)\rightarrow \infty$ (label C). We hence
find that for $\psi(0) > \psi_{\rm min}(0)$ up to three solutions with different masses exist for the same
value of $\psi(0)$. However, when considering the physically relevant parameters $M$ and $Q$ the solutions
are uniquely characterized by these values (see $M$-$Q$-plot in Fig.\ref{fig_2_2}). Again, we find
a ``forbidden band'' of the mass $M$ and charge $Q$ (between label A and label B).  

For $e^2 > e_{\rm cr}^2$ (here $e^2=2$, red lines) we find that only one main branch of fundamental solutions can be constructed.
Contrary to the case with positive $m^2$ this exists for all values of $\psi_0\geq 0$ with the mass $M$ diverging for
$\psi_0\rightarrow \infty$. In this case, solutions exist for all possible values of the mass $M$ and charge $Q$. 

We were also able to construct a branch of excited solutions for $e^2 > e_{\rm cr}^2$ for which the scalar field
function possesses a node. This is indicated by $A'$ in Fig.\ref{fig_2_2}. These solutions exist only
down to a minimal value of the mass $M$ and charge $Q$.

\subsection{The general pattern}
After choosing particular values of the scalar field mass $m^2$ we would now like to
study the general pattern. 
For all values of $e^2$ and fixed, but arbitrary $m^2$ we find that a main branch of solution exists 
that is connected to 
global AdS in the limit $\psi_0 \to 0$. For $e^2 < e^2_{\rm cr}$, the charge $Q$ and mass $M$ of the solutions
on this branch tend to finite values for $\psi_0 \to \infty$ and a second, disjoint branch of solutions
exists. This leads to the observation that in these cases a ``forbidden band'' of the mass $M$ and charge $Q$
exists. On the other hand
for $e^2 > e^2_{\rm cr}$ the mass $M$ and charge $Q$ can have arbitrarily large values in this limit.
We have observed that $e^2_{\rm cr}$ depends on $m^2$ and increases with $m^2$. 
Our numerical results indicate the approximate relation 
\begin{equation}
 e_{\rm cr}^2 \approx  2.4 + \frac{m^2}{3}  \ .
\end{equation}

\subsection{Dependence on $m^2$}
As we have seen above, the $M$-$Q$-plot looks qualitatively similar for all possible values of $m^2$, however
the properties of the solutions given as function of $\psi_0$ depend on $m^2$.
We have hence studied the properties of the solutions in dependence on $m^2$ in detail by fixing $\psi_0$ and $e^2$ and varying
$m^2 \in [-3,1]$.

We find that for those values of $\psi_0$ and $e^2$ for which only one solution exists (which is then a solution
on the branch connected to the AdS vacuum) a variation
of $m^2$ leads to a smooth deformation of the solutions. 
However, we have found it numerically more challenging to vary $m^2$ for those solutions that
are on branches {\it not} connected to the AdS vacuum. The reason is that in this case 
several solutions for one fixed value of $\psi_0$ co-exist.

Let us discuss one example of this type in the following. We have chosen branch $A'$ given in Fig. \ref{fig_2_2}.
This is a branch of first excited solutions possessing one node in the scalar field function.
Moreover, the metric function $f(r)$ possesses a local minimum at say $r = r_{\rm min}$.
We find that fixing $\psi_0$ and  increasing $m^2$
the value of $f(r=r_{\rm min})$ decreases  and tends to zero at some critical value of $m^2=m^2_{\rm cr}$.
This is shown in Fig.\ref{fig_9_bis}, where  we give the metric functions (upper figure) and matter functions (lower figure)
for $e^2=2$, $\psi_0=5.0$ and two different values of $m^2$. 
In this case, we find  $f(r=r_{\rm min}) \approx 0$ for
$m^2\approx 0$. 
At the approach of this critical point we find that $\psi(r > r_{\rm min})\approx 0$ and $a(r> r_{\rm min})\approx 1$ suggesting
that the limiting solution is an extremal Reissner-Nordstr\"om-AdS solution with horizon at $r=r_{\rm min}$. This
is confirmed by the observation that 
the functions $f(r)$ and $\phi(r)$ are given by the analogue expressions for the given values of $M$ and $Q$.  
For $r < r_{\rm min}$ the scalar field remains non-trivial, 
while the electric potential $\phi(r < r_{\rm min})\approx 0$. 
This suggests that the space-time possesses two distinct regions 
matched at $r=r_{\rm min}$: for $r < r_{\rm min}$ the space-time is that of
a compact  uncharged hairy soliton which can be interpreted as an uncharged boson star for $m^2 > 0$, 
while for $r > r_{\rm min}$ the space-time is given by an extremal
Reissner-Nordstr\"om-AdS solution. As a consequence not all of the solutions occurring for negative
values of $m^2$ have a counterpart for $m^2$ positive. 

\begin{figure}
\centering
\epsfysize=17cm
\mbox{\epsffile{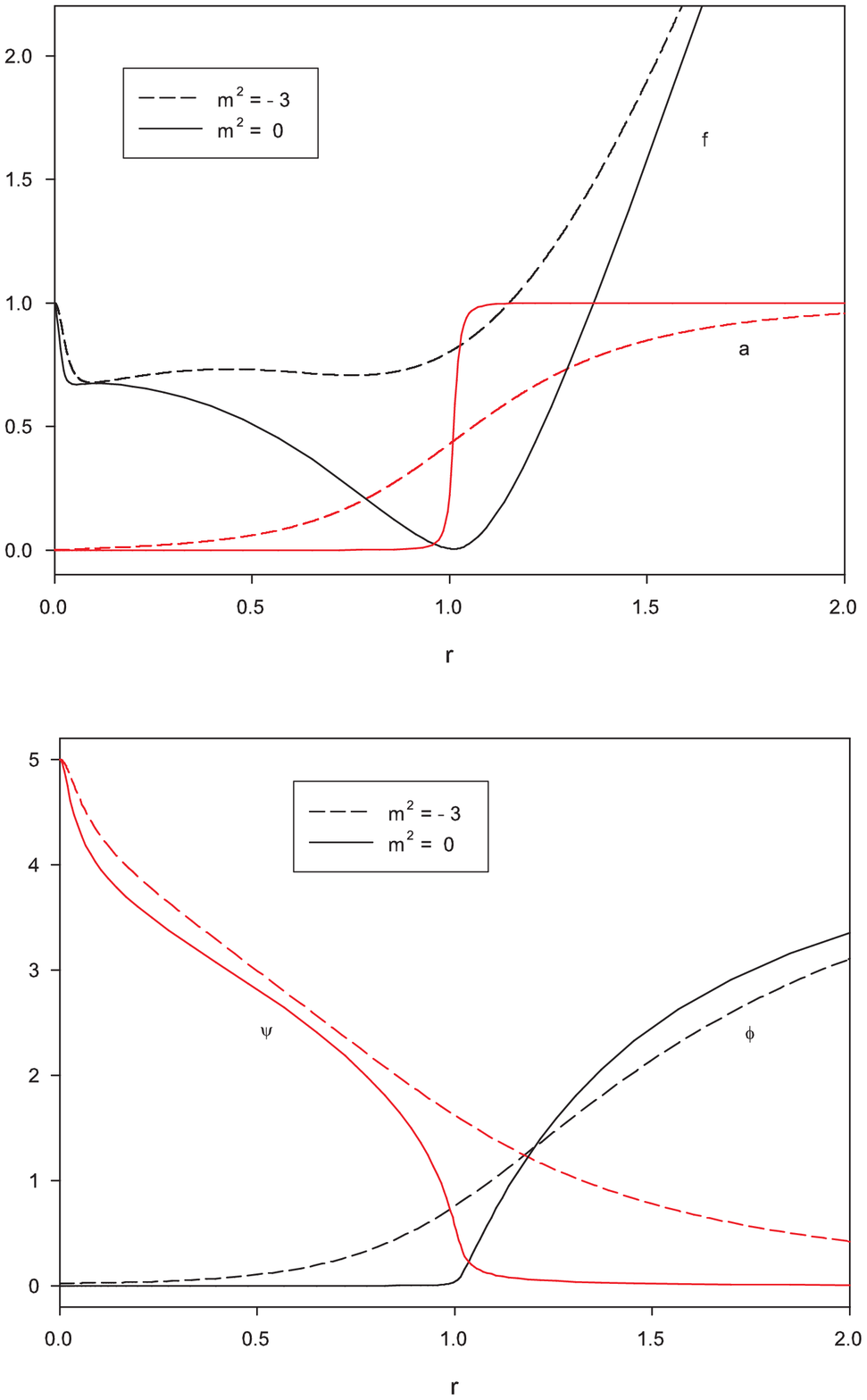}}
\caption{\label{fig_9_bis}
We show (upper figure) the metric functions $f(r)$ (black) and $a(r)$ (red) and (lower figure) the matter functions $\psi(r)$ (red) 
and $\phi(r)$ (black) for $\psi_0=5.0$, $e^2=2$ and
for $m^3= - 3$ (dashed lines) and $m^2=0$ (solid lines), respectively.}
\end{figure}
\begin{figure}
\centering
\epsfysize=8cm
\mbox{\epsffile{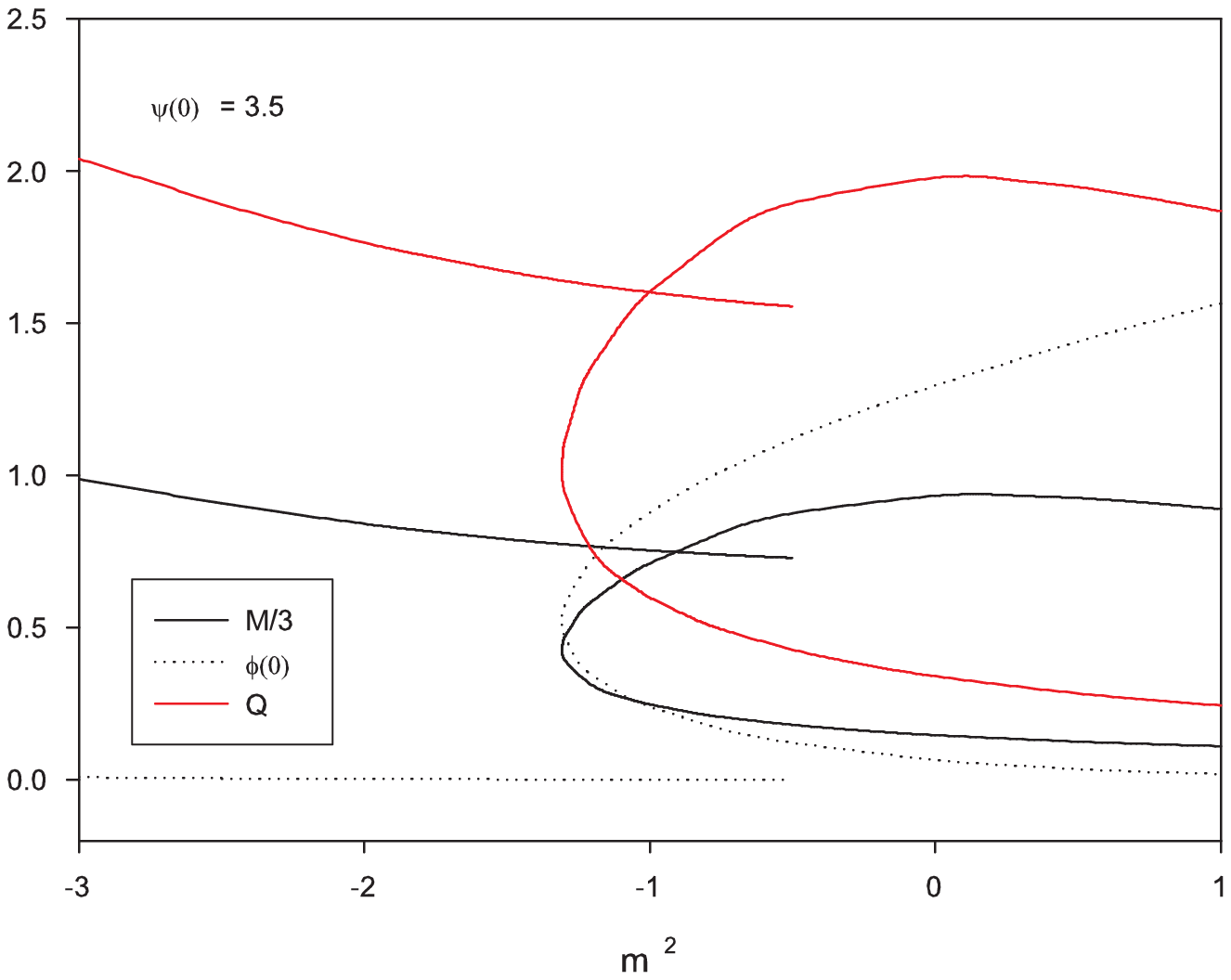}}
\caption{\label{fig_2_m_vary}
We show the mass $M$ (black solid), the value $\phi(0)$ (black dotted) and the charge $Q$ (red solid) of the hairy soliton in dependence on $m^2$ 
for $e^2=2$, $\psi(0)=3.5$.}
\end{figure}

All the phenomena are summarized in Fig.\ref{fig_2_m_vary}, where we show
the mass $M$, the charge $Q$ and
$\phi(0)$ as a function of $m^2$ for $e^2=2$ and $\psi(0)=3.5$. We find that for positive $m^2$ a branch of
solutions exists down to a minimal value of $m^2$ where it bends backwards on a second branch of solutions.
The solutions on this second branch have higher mass $M$ and charge $Q$ as the solutions for the same value
of $m^2$ on the first branch. These are the two disjoint branches labelled A and B in Fig.\ref{fig_2}.
Apparently, these two branches join for sufficiently small $m^2$ such that for $m^2$ small enough
only one branch persists (see Fig.\ref{fig_2_2}).   

Moreover, for negative $m^2$ close to the BF bound only one branch of solutions
exists that extends up to a maximal value of $m^2$. At this value of $m^2$ we find that $\phi(0)=0$ such that
the limiting solution is that described above (see Fig.\ref{fig_9_bis}) consisting of an external extremal Reissner-Nordstr\"om-AdS solution
and an internal uncharged hairy soliton.

\section{Conclusion}
In this paper, we have studied the formation of scalar hair on charged solitons in global AdS space-time.
We have studied the dependence of the solutions on the choice of the charge $e^2$ and the mass $m^2$ of 
the scalar field. For positive $m^2$ we can interpret these solutions as charged boson stars in AdS.
We find that the pattern of solutions depends crucially on the choice of $e^2$ and $m^2$ with
a critical value $e_{\rm cr}^2 \approx 2.4 +m^2/3$ dividing this pattern into distinct types. 
Interestingly, we observe that boson stars in AdS can have arbitrarily large mass and charge, however
also that a ``forbidden band'' of the mass and charge at intermediate values of the mass and charge exists.

In the context of the AdS/CFT correspondence solitons in AdS have been used to describe insulator-superconductor phase transitions
within the condensed matter application \cite{nrt,horowitz_way}, but also to describe 
confinement-deconfinement phase transitions in Quantum
Chromodynamics \cite{witten2,horowitz_myers}. In particular, in \cite{horowitz_way} it has been suggested that
uncharged, planar boson stars in AdS could be the dual of glueball condensates. 
In our case, the boson stars have an asymptotically global AdS and are charged. It would be interesting to understand
what the holographic interpretation of our solutions corresponds to. This is currently under investigation.
\\
\\
{\bf Acknowledgments} B.H. and S.T. 
gratefully acknowledge support within the framework of the DFG Research
Training Group 1620 {\it Models of gravity}. Y.B. would like to thank the Belgian F.N.R.S. for financial
support.

\end{document}